\documentclass[prd,twocolumn,showpacs,preprintnumbers,amsmath,amssymb]{revtex4}
\usepackage{graphicx}% Include figure files
\usepackage{dcolumn}% Align table columns on decimal point
\usepackage{bm}% bold math

\def\be{\begin{eqnarray}}
\def\ee{\end{eqnarray}}
\def\bea{\begin{eqnarray}}
\def\eea{\end{eqnarray}}
\def\kT{{\bf k}_\perp}
\def\qT{{\bf q}_\perp}
\def\rT{{\bf r}_\perp}

\def\bT{{\bf b}_\perp}

\def\0T{{\bf 0}_\perp}
\begin{document}

%\preprint{draft}

\title{Generalized Parton Distributions for large $x$}
% Force line breaks with \\

\author{Matthias Burkardt}
 \affiliation{Department of Physics, New Mexico State University,
Las Cruces, NM 88003-0001, U.S.A.}

\date{\today}% It is always \today, today,

\begin{abstract}
The $t$-dependence of generalized parton distrbutions for
$x\rightarrow 1$ is discussed. We argue that constituent quark 
models, where the $t$-dependence for $x\rightarrow 1$
is through the product $(1-x)t$, are inconsistent. Instead we suggest
a leading dependence in terms of $(1-x)^n t$, where $n\geq 2$,
for $x\rightarrow 1$.
\end{abstract}

\maketitle
\section{Introduction}
Generalized parton distributions (GPDs) \cite{m,ji,r}
are a very powerful 
theoretical
tool which allows linking parton distributions with form factors
as well as many other hadronic matrix elements
(for a recent review, see Ref. \cite{d}).
Unfortunately, they cannot be measured directly but instead they
appear in convolution integrals of the form
\be
\mbox{Amplitude}(\xi,t) \sim \int dx \frac{GPD(x,\xi,t)}{x-\xi \pm i
\varepsilon}. \label{ampl}
\ee
Since these convolution integrals cannot be easily inverted,
GPDs are often `extracted' from the data by writing down a
model ansatz with various free parameters which are
then fitted to the data. In order to reduce the arbitrariness
in this procedure, it is important to incorporate as many
theoretical constraints as possible into the ansatz.
For the intermediate and large $x$ region a commonly used ansatz
for GPDs starts from a simple model for light-cone wave functions.
For example, for the case of the pion one writes down a
2-particle wave function of the form
\be
\psi(x,\kT) \sim f(x)\exp\left( -const.{\cal M}\right),
\label{psi}
\ee
where one conveniently chooses ${\cal M}$ such that wave function
components with a high kinetic energy are suppressed
\be
{\cal M} = \frac{m^2+\kT^2}{x} + \frac{m^2+\kT^2}{1-x} . 
\label{M}
\ee
Upon inserting this type of ansatz into the convolution 
equations for GPDs \cite{overlap} at $\xi=0$ 
\be
H(x,0,t) = \int d^2\kT \psi^*(x,\kT) \psi(x,\kT+(1-x)\qT),
\label{conv}
\ee
one finds \cite{r} a $t$-dependence 
($t\equiv -{\bf q}_\perp^2$)
that is  suppressed by one power of $(1-x)$ for $x\rightarrow 1$
\be
H(x,0,t) = q(x) \exp\left( a t \frac{1-x}{x} \right).
\label{H1}
\ee
Generalizations of Eqs. (\ref{psi}) and (\ref{M}) to more than
two constituents (higher Fock components, baryon) 
yield the same kind of $t$ dependence as Eq. (\ref{H1}).

Obviously, Eq. (\ref{H1}) gives rise to the wrong behavior
for $x\rightarrow 0$ 
(transverse size grows like $\frac{1}{\sqrt{x}}$),
but this does not come as a surprise since one would not expect
a good description  at small $x$ from a valence model.
Moreover, when $Q^2>$ a few $GeV^2$ then the small $x$ behavior  
of GPDs
is practically irrelevant for form factors and Compton scattering.
Therefore, we will not concern ourselves here with the flaws
of the above ansatz at small $x$.

However, it is widely believed \cite{r,all} that Eq. (\ref{H1}) 
provides a 
qualitatively reasonable description in the region of intermediate 
and large $x$, where a constituent model for hadrons has a chance 
to make sense. 

In this letter, we argue that the $x\rightarrow 1$ behavior of 
Eqs. (\ref{psi}-\ref{H1}) is inconsistent.

\section{Transverse Size}
Upon Fourier transforming Eq. (\ref{H1}) to impact parameter
space \cite{s,me:1st,jpr,diehl,ijmpa}, one finds
\be
q(x,\bT) &=& \int \frac{d^2\qT}{(2\pi)^2} e^{i\bT\cdot\qT}
H(x,0,-{\bf q_\perp^2})\nonumber\\
&=& q(x) \frac{1}{4\pi a} \frac{x}{1-x}
\exp\left( -\frac{{\bf b_\perp^2}}{4a} \frac{x}{1-x} \right),
\ee
i.e. the width of this distribution in impact parameter space
behaves like 
\be
\langle \bT^2 \rangle \equiv 
\frac{\int d^2\bT \bT^2 q(x,\bT)}{\int d^2\bT q(x,\bT)}
\propto {1-x}
\ee
as $x\rightarrow 1$. However,  $\bT$ only measures the distance
from the active quark to the center of momentum of the hadron.
A better measure for the size (diameter) of a configuration
of the wave function is given by
the separation $\rT$ between the active quark and the center of 
momentum of the spectators
(actually even $\rT$ is only a lower bound on the diameter). 
In terms of $\bT$ one finds for the
separation $\rT=\frac{\bT}{1-x}$. 
With the above ansatz (\ref{H1}) the transverse size  diverges as $x \rightarrow 1$
\be
\langle \rT^2 \rangle \equiv 
\frac{\int d^2\bT \frac{\bT^2}{(1-x)^2} q(x,\bT)}
{\int d^2\bT q(x,\bT)}
\sim \frac{1}{{1-x}}.
\label{r}
\ee
For this result it was irrelevant that the $t$-dependence in
Eq. (\ref{H1}) is exponential. The important feature that led to
Eq. (\ref{r}) was the fact that the dependence on $t$ was through
the combination $(1-x)t$ for $x\rightarrow 1$.

This power law growth of the transverse size as $x\rightarrow 1$ 
is not only bizarre,
but in fact it makes the logic behind the valence ansatz
(\ref{psi}), (\ref{M}) for the light-cone wave function 
inconsistent: First of all, 
if a $q\bar{q}$ pair is separated by a large $\perp$
distance then its potential energy is very high and therefore one
cannot neglect the potential energy in Eq. (\ref{M}).
In the next section we will provide an estimate of that potential
energy.
Secondly, a $q\bar{q}$ pair that has a $\perp$ separation must
be connected by a $\perp$ gauge string --- otherwise its
energy is infrared divergent. Even if one does not put in such a
gauge string ``by hand'', any nonperturbative diagonalization
of a light-cone Hamiltonian for QCD should yield such a string
in the wave function for finite energy hadrons. The presence of
such a $\perp$ gauge string automatically implies that this component
of the wave function also contains gluon degrees of freedom
in contradiction with the valence ansatz that was used as
a starting point (\ref{psi}) and (\ref{M}). In fact, since the
$\perp$ separation between the quark and the antiquark diverges
as $x\rightarrow 1$, such a state would have to contain an infinite
number of gluons as $x\rightarrow 1$.
Similar reasoning applies to a nucleon valence ansatz analogous to
Eq. (\ref{M}).

In summary, the whole logic that starts from a valence ansatz for the 
light-cone wave function, in which the $\perp$ momentum dependence is
only governed by the kinetic energy, becomes inconsistent in the limit
$x\rightarrow 1$. The above light-cone wave function model
is the only justification for writing down a $t$-dependence of the
form $\exp \left(at\frac{1-x}{x}\right)$ for large $x$. Hence we are
led to abandon Eq. (\ref{H1}) for large $x$.
Of course, for intermediate $x$, Eq. (\ref{H1})
may still provide a reasonable and consistent description. 
However, since form factors and Compton amplitudes, 
are rather sensitive to the behavior of $H(x,0,Q^2)$ for $x>0.5$ when 
$Q^2 > 10 GeV^2$, it becomes necessary to improve on
the $x\rightarrow 1$ bahavior of Eq. (\ref{H1}).

\section{An Improved Ansatz for $H(x,0,t)$}

If one really wants to know $H(x,0,t)$ for $x\rightarrow 1$, all 
one has to do is solve QCD. Since we are not yet ready to perform
such calculations with the required accuracy, we want to propose in 
this section an improved model ansatz for $H(x,0,t)$ for 
$x\rightarrow 1$.

Even without doing any calculation, it is clear that in order to cure
the problem of increasing size as $x\rightarrow 1$, the 
$t$-dependence must be suppressed with a higher power of $(1-x)$:
a finite $\perp$ size as $x\rightarrow 1$ is
achieved if and only if the dependence on $t$ for $x\rightarrow 1$
is of the form $t (1-x)^n$ with $n\geq 2$. In this section, we
would like to present a variational argument to provide a more precise
estimate for the $x\rightarrow 1$  behavior.

For this purpose, let us estimate the potential energy contribution
to the light-cone Hamiltonian for a $q\bar{q}$ pair that is 
separated by a $\perp$ distance $\rT$.
Assuming a linear potential, the effective mass of the QCD string
connecting the $q\bar{q}$ pair is at least 
\be
m_g = \sigma \left|\rT\right|.
\ee
The quantity that enters the light-cone Hamiltonian is the invariant
mass of the glue divided by the light-cone momentum carried by the
glue. Without actually solving (i.e. diagonalizing) the light-cone 
Hamiltonian one cannot know how the light-cone momentum is divided
among the antiquark and the string of glue, but obviously the
glue cannot carry more than momentum fraction $1-x$ if  $x$ is the
momentum fraction carried by the active quark. This motivates us to 
consider the following (conservative) ansatz
for the light-cone energy of the $q\bar{q}$ pair including the effects
of the potential energy at large separations
\be
\tilde{\cal M} = {\cal M} + \frac{\sigma^2 \rT^2}{1-x},
\label{M2}
\ee
where  $\sigma \approx (440\,MeV)^2$ is the string tension.
This ansatz is conservative because, as we explained above, it 
only underestimates the light-cone energy of the glue. Nevertheless,
let us estimate the effect of adding such a term in the effective
Hamiltonian for a constituent model of the pion.

For $x\rightarrow 1$, the variables $\kT^2$ and $\rT^2$ (which are
Fourier conjugate to each other) appear symmetrically in 
$\tilde{\cal M}$ (that is up to the factor $c\sigma^2$, which 
provides the scale): both are divided by one power of $(1-x)$.
Therefore if one considers  in the light-cone energy an ansatz that
includes the transverse gauge string tension
(in this ansatz we are only concerned about the singularity in the
energy for $x\rightarrow 1$)
\be
{\tilde{\cal M}} = \frac{m^2}{x(1-x)} + \frac{\kT^2}{x(1-x)}
+ \frac{\sigma^2\rT^2}{1-x}
\ee
one expects that the $\kT$-dependence of the 
light-cone wave function factorizes near $x\rightarrow 1$, i.e.
\be
\psi(x,\kT)\stackrel{x\rightarrow 1}{\longrightarrow}
(1-x)^k \phi(\kT).\label{factor}
\ee
Inserting this result into the convolution formula (\ref{conv})
yields GPDs where the $t$-dependence is suppressed by
2 powers of $(1-x)$ near $x\rightarrow 1$.

One may criticize that instead of solving the
light-cone Hamiltonian for QCD, we only performed some dimensional
analysis. However, there is certainly no justification to 
suppress in an ansatz of the light-cone wave function only 
components that have a high kinetic energy but not those that
have a high potential energy (\ref{psi}) --- especially
if the resulting state yields a potential energy that diverges badly
as $x\rightarrow 1$. In contrast, our ansatz at least builts in
some of the effects from the gluon field and thus gives rise to a
finite size for $x\rightarrow 1$, and
potential and kinetic energy
scale in the same way for $x\rightarrow 1$. A finite size
in position space suggests a finite size in momentum space. Since
the convolution expressions for GPDs involve the $\perp$ momentum
transfer with a factor of $(1-x)$ this naturally leads to
GPDs where the $t$-dependence for $x\rightarrow 1$ should be
through the combination
$t(1-x)^2$, i.e. a better ansatz to describe the $x\rightarrow 1$
behaviour of GPDs reads
\be
H(x,0,t) \stackrel{x\rightarrow 1}{\longrightarrow}  
q(x) \exp \left[at (1-x)^2\right]. \label{2}
\ee
However, we neglected many things in our analysis. For example,
the ansatz for the potential energy term included on the
$\perp$ string tension and we potentially we even underestimated 
that contribution, which leaves open the possibility that the
suppression of the $t$-dependence may be of the form $t(1-x)^n$
with $n\geq 2$. Although our analysis clearly rules out $n<2$,
it is too crude to specify the correct value of $n$ uniquely.

Of course, it is quite possible that 
the actual behavior of GPDs at large $x$ is even more
complicated than the semi-factorized form in Eq. (\ref{2}). 
Because of this possibility, one should regard Eq. (\ref{2}) only
as one possibility to illustrate the difference to the previously
used ansatz. Nevertheless, our main point, i.e. the fact that the 
$t$-dependence of GPDs for $x\rightarrow 1$ should be suppressed by
at least 2 powers of $(1-x)$ should be model-independent. This
general result is also supported by perturbative QCD \cite{yuan} 
studies as well as recent lattice gauge theory results \cite{dru}. 
In Ref. \cite{yuan} it was found that in pQCD the $t$-dependent 
terms near
$x\rightarrow 1$ are suppressed by an additional power of $(1-x)^2$
near $x\rightarrow 1$. Lattice gauge theory calculations
of the r.m.s. radii for the lowest moments of $H(x,0,t)$ indicate
a strong suppression for the r.m.s. radii of subsequent moments,
which indicates a suppression of the $t$ dependence with $(1-x)^n$
near $x\rightarrow 1$ where $n>1$. Recent transverse lattice
calculations \cite{dalley} even indicate a shrinking $\perp$ size
as $x\rightarrow 1$, i.e. $n>2$, in the case of the pion.

It is interesting to note that GPDs, which have a $t$-dependence 
that comes with a factor of $(1-x)^2$ near $x\rightarrow 1$,
naturally give rise to
Drell-Yan-West duality between parton distributions at large $x$
and the form factor. Indeed, the ansatz
\be
H(x,0,t)=(1-x)^{2N_s-1} \exp\left[at(1-x)^2\right]
\label{DYW}
\ee
yields
\be
F(t)&&= \int dx H(x,0,t)\\
&&\stackrel{t\rightarrow - \infty}{\longrightarrow} 
\frac{\Gamma (N_s)}{2a^{N_s}}
\frac{1}{(-t)^{N_s}}.
\ee
Here only the behavior near $x\rightarrow 1$ matters and
DYW duality would also arise (with a different coefficient) if
the exponential function were being replaced by any function that
falls rapidly for $t\rightarrow -\infty$ , $x$ fixed.

We should also point out that there is an interesting connection
between the value of $n$ and the occurrence of color transparency
\cite{miller}, where color transparency does not occur for
$n<2$.

\section{Summary}
We have demonstrated that a commonly employed ansatz for light-cone 
wave functions,
where the $\kT$-dependence is only through the light-cone kinetic 
energy of the constituents, is inconsistent for $x\rightarrow 1$ 
because it leads to a divergent transverse size for those wave
function components where one constituent carries $x\rightarrow 1$.
If the constituents are separated by a divergent $\perp$ distances
then they must be connected by $\perp$ strings of color flux
of infinite length, which
in the infinite momentum frame correspond to an infinite number
of transverse gluons. 
Therefore such a configuration corresponds to a very high Fock 
component and {\sl not} to a valence component. Hence the initial
valence ansatz becomes inconsistent and there is no reason to 
trust the $x\rightarrow 1$ behavior of GPDs that are based on 
such a valence
ansatz (with $\kT$ dependence in the wave function only through
the light-cone kinetic energy). Since such an ansatz has been the
only motivation for parameterizing GPDs with function where the
leading $t$ dependence for $x\rightarrow 1$ is through the product
$(1-x)t$, we are led to abandon such parameterizations for GPDs.

The origin for this divergent size problem is the fact that the 
distance between the active quark and the center of momentum of the
spectators ${\bf r}\equiv {\bf r_{\perp q}}-{\bf r_{\perp\bar{q}}}$
is related to the impact parameter $\bT$ 
(the distance from the active
quark to the center of momentum) via
$\rT = \bT/(1-x)$. In order for the hadron to have a finite size
as $x\rightarrow 1$, one must have $\langle \bT^2\rangle \sim
(1-x)^n$ with $n\geq 2$, which in turn requires that the leading
dependence on $t$ is through the product $(1-x)^n t$ with $n\geq 2$.

While we are not able to predict what the actual behavior is
for $x\rightarrow 1$, we made an attempt to include the gluonic
energy into an ansatz for the light-cone wave function. With such
an ansatz, large size configurations are naturally suppressed and
we find GPDs where the $x\rightarrow 1$ dependence on $t$ is
through the combination $(1-x)^n t$ with $n=2$. Of course the
estimate that led to this result was very crude and therefore one 
should not take the value $n=2$ as a rigorous prediction. 
Nevertheless, we believe that $n\geq 2$ 
is a much more reasonable choice
in parameterizations than $n=1$ since $n=2$ is consistent
with hadrons that
have a finite size (actually for $n>2$ it would even be consistent
with a vanishing size) for $x\rightarrow 1$.

Another interesting observation concerns the Drell-Yan-West
duality between the form factor and structure functions since
the case $n=2$ naturally leads to the same duality relation as
quark counting rules.

{\bf Acknowledgements:}
I would like to thank G. Miller and D. Renner for stimulating
discussions.
This work was supported by the DOE under grant number 
DE-FG03-95ER40965.

\bibliography{largeX.bbl}% Produces the bibliography via BibTeX.
\end{document}